\documentclass[conference]{IEEEtran}
\IEEEoverridecommandlockouts
\usepackage{cite}
\usepackage{amsmath,amssymb,amsfonts}
\usepackage{algorithm}
\usepackage{graphicx}
\usepackage{textcomp}
\usepackage{xcolor}
\usepackage{tabularx}
\usepackage{booktabs}
\usepackage{subfigure}
\usepackage{hyperref}
\def\BibTeX{{\rm B\kern-.05em{\sc i\kern-.025em b}\kern-.08em
		T\kern-.1667em\lower.7ex\hbox{E}\kern-.125emX}}
\begin{document}
	
	\title{Blind Estimation of Room Acoustic Parameters and Speech Transmission Index using MTF-based CNNs\\
		\thanks{This work is supported by Shibuya Science Culture and Sports Foundation. This work was supported by JSPS-NSFC
			Bilateral Joint Research Projects/Seminars
			(JSJSBP120197416), SCOPE Program of Ministry of Internal Affairs and Communications (Grant Number: 201605002). This work is also supported by Thammasat University Basic and Applied Research Grant.}
	}
	
	\author{
		\IEEEauthorblockN{Suradej Duangpummet\IEEEauthorrefmark{1}\IEEEauthorrefmark{2}\IEEEauthorrefmark{3}, Jessada Karnjana\IEEEauthorrefmark{2}, Waree Kongprawechnon\IEEEauthorrefmark{3}, and Masashi Unoki\IEEEauthorrefmark{1}}\\
		
		\IEEEauthorblockA{\IEEEauthorrefmark{1}  \textit{School of Information Science, Japan Advanced Institute of Science and Technology, Japan}\\
		}
		
		\IEEEauthorblockA{\IEEEauthorrefmark{2} \textit{National Science and Technology Development Agency,  Thailand}\\
		}
		
		\IEEEauthorblockA{\IEEEauthorrefmark{3}   \textit{Sirindhorn International Institute of Technology, Thammasat University, Thailand}\\
		}
		
		\IEEEauthorrefmark{1}\{suradej, unoki\}@jaist.ac.jp, \IEEEauthorrefmark{2}jessada.karnjana@nectec.or.th, \IEEEauthorrefmark{3}waree@siit.tu.ac.th
	}
	\maketitle
	
	\begin{abstract}
		Room acoustic parameters, such as reverberation time ($T_{60}$) and clarity ($C_{80}$), as well as an objective index, the speech transmission index (STI), are essential in acoustics. Such parameters and STI are, however, difficult to obtain in everyday places with people. Blind estimation without measuring room impulse response (RIR) is necessary and challenging. This paper proposes a method based on the modulation transfer function (MTF) and Schroeder's RIR model for estimating $T_{60}$s in seven-octave bands. 
		The estimated $T_{60}$s are used to approximate the MTF and RIR. The STI and five room-acoustic parameters, including $T_{60}$, early decay time (EDT), $C_{80}$, Deutlichkeit ($D_{50}$), and center time ($T_s$), can therefore be estimated. Convolutional neural networks (CNNs) were used for mapping temporal amplitude envelopes of a reverberant speech signal to $T_{60}$s for the sub-bands. Simulations were carried out by estimating the five parameters and STI from unseen reverberant speech signals. The root-mean-square errors between ground-truths and estimated parameters suggest that the accuracy of estimated $T_{60}$ and STIs can be improved by about $40\%$ and $25\%$ compared with previous methods, respectively. The other parameters were also correctly estimated, and they are comparable with those obtained from standard measurements.
	\end{abstract}
	
	
	\section{Introduction}
	Subjective aspects in speech and music assessments, such as speech intelligibility and music clarity in enclosures, can be objectively described through room acoustic parameters and objective indices \cite{Kuttruff}. Room acoustic parameters are also crucial for architectures or acousticians who are involved in an auditorium \cite{AIJ}. Most of the parameters in ISO $3381$, such as reverberation time ($T_{60}$), early decay time (EDT), clarity ($C_{80}$), Deutlichkeit ($D_{50}$), and center time ($T_s$), are derived from room impulse response (RIR) \cite{ISO3381}. Similarly, a speech transmission index (STI), which is an objective index in IEC $60268$, can be calculated by measuring the modulation transfer function (MTF) or can be derived from the RIR \cite{Houtgast_Steeneken_MTF,IEC60268_STI}. Therefore, the RIR or MTF needs to be measured in general.
	
	However, it is difficult to measure RIR or MTF in daily places where people cannot be excluded (e.g., stations, airports, and schools). As a result, many methods have been proposed to estimate such a parameter without measuring the RIR as so-called blind estimation. Unoki {\it et al.} proposed methods based on the concept of the MTF to estimate $T_{60}$ and STI \cite{MU_T60,MU_STI_EUSIPCO13, MU_STI}. Kendrick {\it et al.} proposed the maximum likelihood estimation to approximate energy decay curves from reverberant speech and music \cite{Kendrick}. The energy decay curve is used for calculating $T_{60}$ and EDT, as described in ISO $3381$. Techniques based on deep neural networks (DNNs) were also successful, such as a deep convolutional neural network (CNN) for estimating STI \cite{Prem}. For $T_{60}$ estimation, many approaches have been evaluated in the Acoustic Characterization of Environments (ACE) Challenge \cite{ACE}, for example, the CNN with spectra-temporal features in the time-frequency domain \cite{CNN_T60}. Also, a recent combination of a CNN and long short-term memory (LSTM) network has been proposed \cite{deng2020IS}. 
	
	In addition to STI and $T_{60}$, Parada {\it et al.} proposed an estimator for the clarity index at $50$ ms ($C_{50}$) by using a spectral envelope in the modulation-domain with a bidirectional LSTM \cite{Parada}. We previously proposed a robust method to estimate STI by using the full-band temporal amplitude envelope (TAE) of a noisy reverberant speech signal with a CNN \cite{Me_APSIPA2019}. This method could overcome a mismatch problem between the model and real acoustic conditions.
	
	However, current blind estimation methods, as the aforementioned ones, can estimate only a single parameter. It is limited to a specific aspect and is inadequate to describe the characteristics of room acoustics completely. To this end, in this paper, we propose a scheme to simultaneously estimate multiple parameters based on the basis of the MTF, Schroeder's RIR model, and CNNs for sub-bands. Therefore, $T_{60}$, EDT, $C_{80}$, $D_{50}$, $T_s$, and STI can be simultaneously estimated from a speech signal in reverberant environments.
	\begin{figure*}[ht]
		\centering
		\includegraphics[width= 1.0 \textwidth]{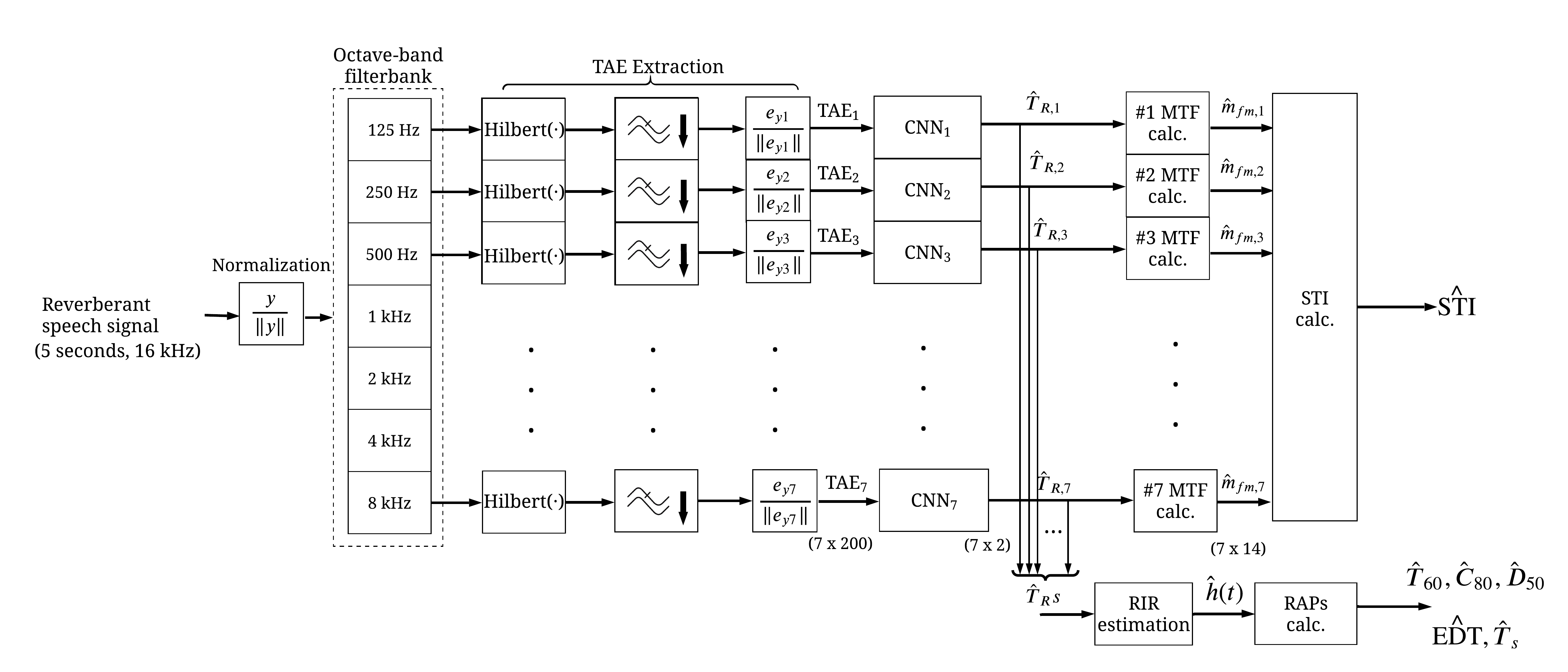}
		\caption{Block diagram of proposed method.}
		\label{fig:ProposedMethod}
	\end{figure*}
	
	\section{Proposed Method}
	We propose a scheme for estimating five room-acoustic parameters and an STI, namely MTF-based CNNs, as shown in Fig.\,\ref{fig:ProposedMethod}. The scheme incorporates the MTF concept into a nonlinear regression using CNNs. The $T_{60}$s for sub-bands are mapped in accordance with the characteristics of the TAEs under reverberant conditions. RIR is approximated from the estimated $T_{60}$s to derive the five parameters and STI.  
	
	\subsection{Definitions}
	In a reverberant environment, we assume that an observed signal, $y(t)$, is the result of the convolution between an original speech, $x(t)$, and RIR, $h(t)$. The RIR is used to represent acoustic characteristics of a given room in the time domain. In the modulation-frequency domain, the MTF is used to quantify the effect of reverberation \cite{Houtgast_Steeneken_MTF}. The MTF is defined as
	\begin{equation}
		m(f_m) = \displaystyle\frac{\displaystyle\int_{0}^{\infty} h^2(t)e^{-j2\pi{f_m}t}dt}{\displaystyle\int_{0}^{\infty} h^2(t)dt},
		\label{eq:MTF}
	\end{equation}
	where $m(f_m)$ is the MTF at a modulation frequency, $f_m$. In this study, $h(t)$ is the RIR model proposed by M. R. Schroeder \cite{SchroederRIR}, and it is defined as
	\begin{equation}
		h(t)\!=\!e_h(t)\,c_h(t)\!=\! a\,\text{exp}\left(-\frac{6.9t}{T_{60}}\right)\,c_h(t),
		\label{eq:schroederRIR}
	\end{equation}
	where $e_h(t)$ is an exponential decay, $c_h(t)$ is a carrier of white Gaussian noise (WGN), and $a$ is a gain factor. The MTF according to Schroeder's RIR can be expressed as
	\begin{equation}
		m(f_m, T_{{60}})= {\left[ 1+\left(2\pi f_m \frac{{T_{{60}}}}{13.8}\right)^2\right]}^{-{\frac{1}{2}}}.
		\label{eq:mtf_T60_k}
	\end{equation}
	According to ISO\,\,$3381$ and IEC\,\,$60268$, the definitions of the interested room acoustic parameters and STI are as follows. $T_{60}$ is the period in seconds unit from the energy decay curve of the RIR when the curve decreases by $60$ dB. The period of the energy decay curve by $10$ dB is the EDT.
	
	$C_{80}$ and $D_{50}$ are the energy ratio between the reflection components and total energy of the RIR. $C_{80}$ is used to characterize the transparency of music halls in dB unit and is defined as
	\begin{equation}
		C_{80}\!=\!10 \log_{10}\frac{\displaystyle\int_{0}^{80\rm{ms}} {h^2(t)} dt}{\displaystyle\int_{80\rm{ms}}^{\infty} { h^2(t)} dt}.
		\label{eq:C80}
	\end{equation}
	
	$D_{50}$ is used to evaluate the speech intelligibility of lecture halls and classrooms in percent and is defined as
	\begin{equation}
		D_{50}\!=\!\frac{\displaystyle\int_{0}^{\rm{50ms}} {h^2(t)} dt}{\displaystyle\int_{0}^{\infty} { h^2(t)} dt}\times 100.
		\label{eq:D50}
	\end{equation}
	
	Center time, $T_s$, is the period at the center of gravity of the RIR. $T_s$ shows the balance between clarity and reverberation related to speech intelligibility and is defined as
	\begin{equation}
		T_s\!=\!\frac{\displaystyle\int_{0}^{\infty} { h^2(t)}\cdot t\, dt}{\displaystyle\int_{0}^{\infty} {h^2(t)}dt}.
		\label{eq:Centre time}
	\end{equation}
	
	Finally, the STI, an objective index, is used to assess the speech transmission quality from a talker to a listener of a given room \cite{IEC60268_STI,Houtgast_Steeneken_MTF}. Hence, speech intelligibility can be predicted by calculating STI in a scale from $0$ to $1$. The STI algorithm is based on the measurement of the MTF in sub-bands. $98$ modulated stimuli are used to calculate the distortion ratios between the inputs and observed signals. The stimuli are amplitude-modulated signals from seven-octave-band carriers and $14$ modulation frequencies. The STI is calculated by weighting the modulation transmission indices of the seven-octave bands. See \cite{IEC60268_STI} for more details. However, since the MTF can be derived from RIR, as shown in (\ref{eq:MTF}) and known as the indirect method, the STI can be calculated from the RIR.  
	
	\subsection{Sub-band analysis}
	The sub-band analysis for estimating room acoustic parameters is derived from the STI algorithm, which is from the basis of the MTFs in seven-octave bands. Thus, we exploit the relation between the MTF and RIR, as shown in (\ref{eq:MTF}), within the same bands as the STI. The bands have center frequencies ranging from $125$ Hz to $8$ kHz. The normalized reverberant-speech signal is the input. The signal is then decomposed to each sub-band using octave-band filters.
	
	Based on the MTF concept, a temporal envelope of any signal is a smoothed version of the original signal when it is passed through a reverberant room \cite{Houtgast_Steeneken_MTF}. We then utilize the seven TAEs to represent the modulation distortion characteristics caused by reverberation in the bands. The reverberation, in terms of the $T_{60}$s, attenuates the observed TAEs. The seven TAEs account for the accuracy enhancement of the estimating $T_{60}$ and STI as well as the other parameters. 
	
	The TAE in each band is extracted according to (\ref{eq:TAE_extraction}). The observed signal is decomposed by using the Hilbert transform and a lowpass filter (LPF). The LPF is a sixth-order Butterworth filter with a cut-off frequency of $20$ Hz. We downsample the signal to $40$ Hz to reduce the computation complexity. Then, the TAEs are mapped to their associated $T_{60}$s for the seven-octave bands by using CNNs.
	\begin{equation}
		{{e}_y(t)}\!=\!\rm{LPF}\,[|\textit{y}(\textit{t})+\textit{j}\cdot{\rm{Hilbert}}(\textit{y}(\textit{t}))|]\,.  
		\label{eq:TAE_extraction}
	\end{equation} 
	\subsection{MTF-based CNN models}
	We use seven one-dimensional CNNs for mapping the characteristics of the TAEs with their associated $T_{60}$s. There are seven identical models for each band. Each model consists of four convolutional layers. The input layer takes a TAE to be convoluted with the filters. The regulated linear unit (ReLU), $f(x)\!=\!\max(x,0)$, performs nonlinear activation in every convolutional layer. Batch normalization is applied after the first convolution. Max pooling is also used for reducing the dimensions before the next layer. The dropout rate before the last layer is set to $20\%$ to avoid the memorized problem for a number of dominant nodes. The fully connected layer is the output layer. The seven CNNs are trained from the TAEs/$T_{60}$s pairs. The trained networks are supervised by the $T_{60}$s' ground-truths. The ground-truths are calculated from simulated RIRs. The simulated RIRs are synthesized by using Schroeder's RIR model, as in (\ref{eq:schroederRIR}). The output of each CNN for each sub-band is the estimated $T_{60}$. The details of the MTF-based CNN model is shown in Table. \ref{tab:Network}.
	
	\begin{table}[ht]
		\centering
		\caption{Network architecture of the MTF-based CNN model.}
		\label{tab:Network}
		\begin{tabular}{@{}cll@{}}
			\toprule
			No. & \multicolumn{1}{c}{Layer Type} & \multicolumn{1}{c}{Parameters}    \\ \midrule
			$1$   & Input                          & TAE shape$\,=\,1\times200$                             \\
			$2$   &$\text{Conv1D}^{\small{1}\text{st}}$                     &$32$ filters, filter size $\!=\!10\times1$, ReLU \\
			$3$  & Pooling                    &  $\text{max pooling, size}=2, \text{stride}\,=\,1$                        \\	
			$4$  &$\text{Conv1D}^{\small{2}\text{nd}}$                    &$ 16$ filters, filter size $\!=\!5\times1$, ReLU \\
			$5$  & Pooling                    &  $\text{max pooling, size}=2, \text{stride}\,=\,1$                        \\
			$6$   & Dropout                        & $0.2$                               \\
			$7$  &$\text{Conv1D}^{\small{3}\text{rd}}$              &$8$ filters, filter size $\!=\!5\times1$, ReLU \\
			$8$  & Pooling                    &  $\text{max pooling, size}=2,$\\
			$9$  &$\text{Conv1D}^{\small{4}\text{th}}$              &$4$ filters, filter size   $\!=\!5\times1$, ReLU                   \\
			$10$  & Fully Connected                & $1$ output (i.e., $T_{60}$), ReLU                            \\
			$11$  & Regression Output              & mean-square-error (MSE)      \\ \bottomrule  	
		\end{tabular}
	\end{table}
	
	\begin{figure}[h]
		\centering
		\includegraphics[width=0.45\textwidth]{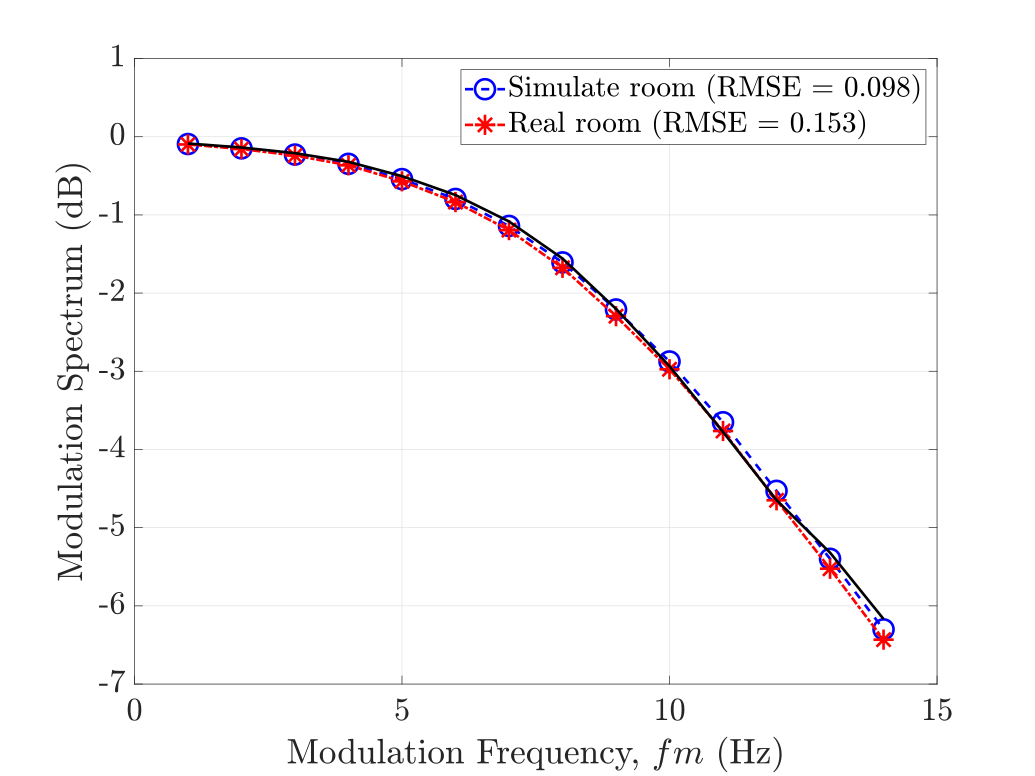}
		\caption{Example of the approximated MTFs from the estimated $T_{60}$s, where dashed lines are the estimated MTFs, and the solid line is the ground-truths.} 
		\label{fig:estMTF_fromT60s}
	\end{figure}
	\subsection{RIR approximation}
	The estimated $T_{60}$s are used to approximate RIR, $\hat{h}(t)$. The approximated RIR is reconstructed by using the Schroeder's RIR model. The Schroder's RIR depends on only the reverberation time of a room. Hence, the estimated $T_{60}$ for each octave-band can construct the temporal envelope of the RIR, $\hat{e_h}(t)$. The envelope is modulated with a carrier signal with is the band-limited Gaussian noise with a bandwidth of $1/3$ of an octave. The sub-band RIRs are then summed together to be the approximated RIR. The RIR reconstruction is defined as 
	\begin{equation}
		\hat{h}(t)\!=\!\sum_{k=1}^{K} \text{exp} \left( -\frac{6.9t}{{T_{60,k}}} \right) c_{h,k}(t),
		\label{eq:ReconRIR}
	\end{equation}
	where ${T_{60,k}}$ is the estimated $T_{60}$ in the $k$-th band and $K=7$, and $c_{h,k}(t)$ is band-limited Gaussian noise. The STI can then be calculated from the estimated $T_{60}$s based on the basis of the MTF, as shown in (\ref{eq:mtf_T60_k}). Also, the $T_{60}$, EDT, $C_{80}$, $D_{50}$, and $T_s$ can be calculated by following the definitions from (\ref{eq:C80}), (\ref{eq:D50}), and (\ref{eq:Centre time}), respectively.
	
	\begin{figure*}[htpb!]
		\hfill
		\begin{subfigure}
			\centering
			\includegraphics[width=0.468\textwidth]{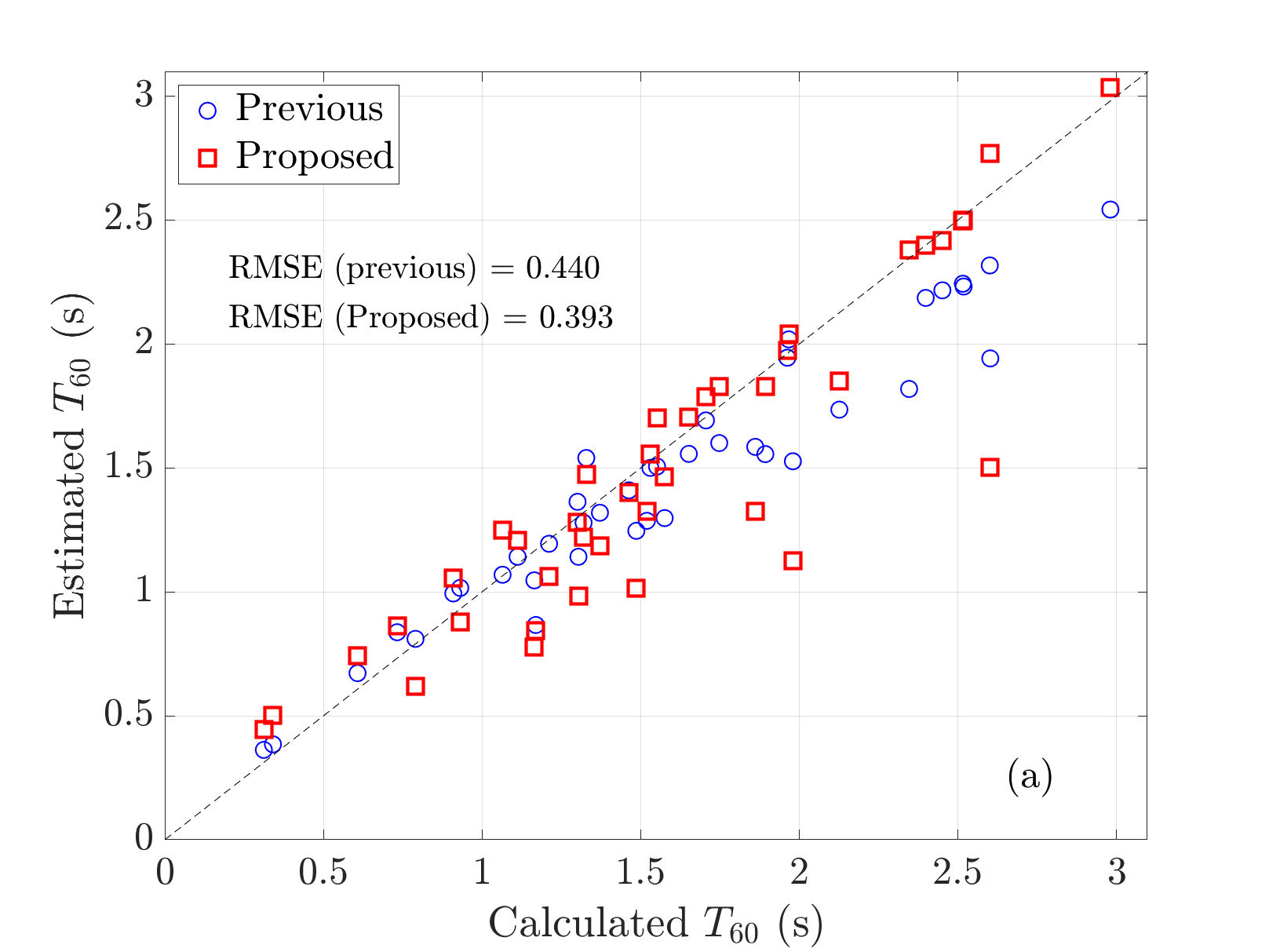}
			\label{fig:T60_SMILE}
		\end{subfigure}
		\hfill
		\begin{subfigure}
			\centering
			\includegraphics[width=0.49\textwidth]{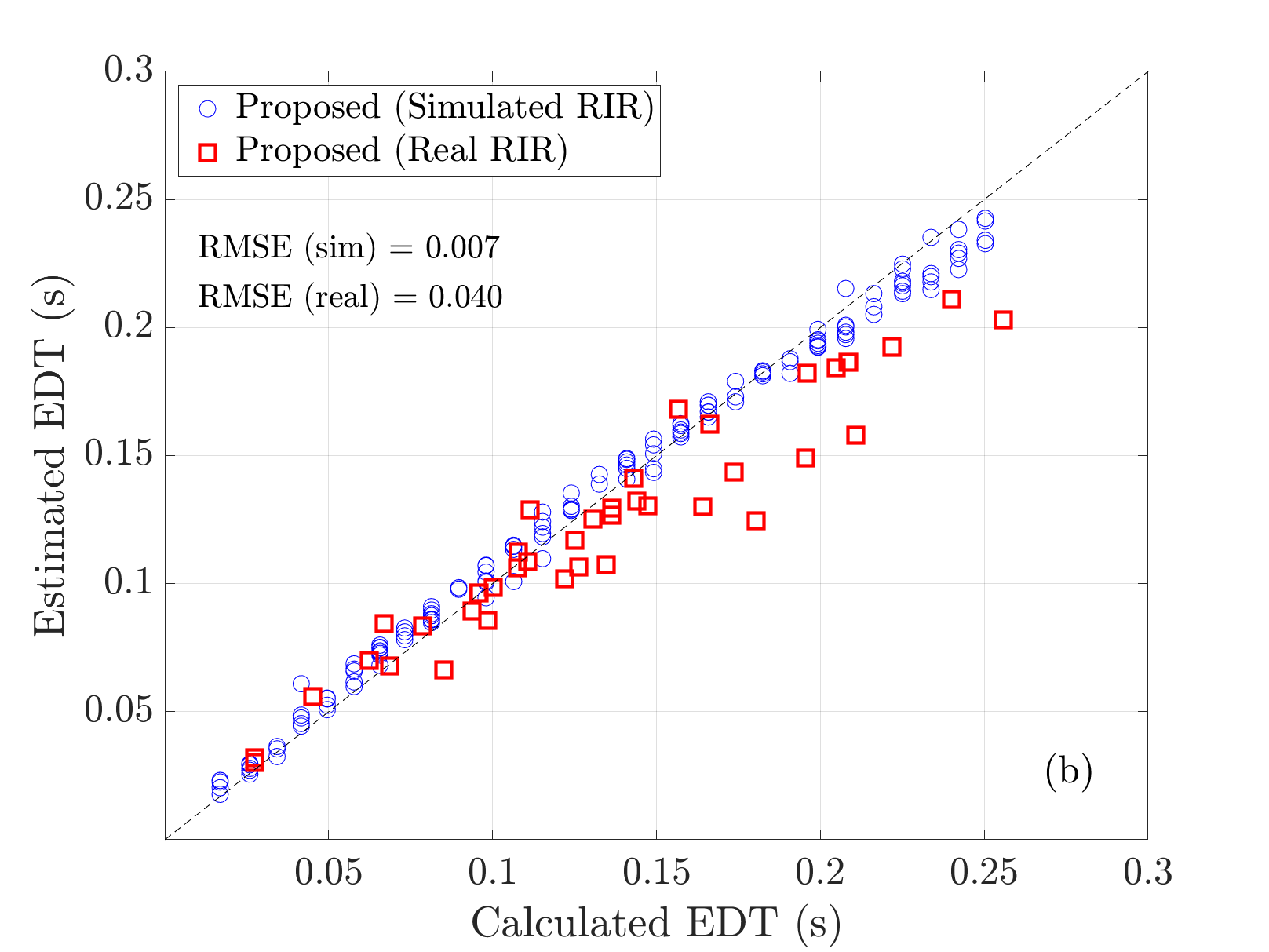}
			\label{fig:EDT_SMILE}
		\end{subfigure}
		
		\hfill
		\begin{subfigure}
			\centering
			\includegraphics[width=0.48\textwidth]{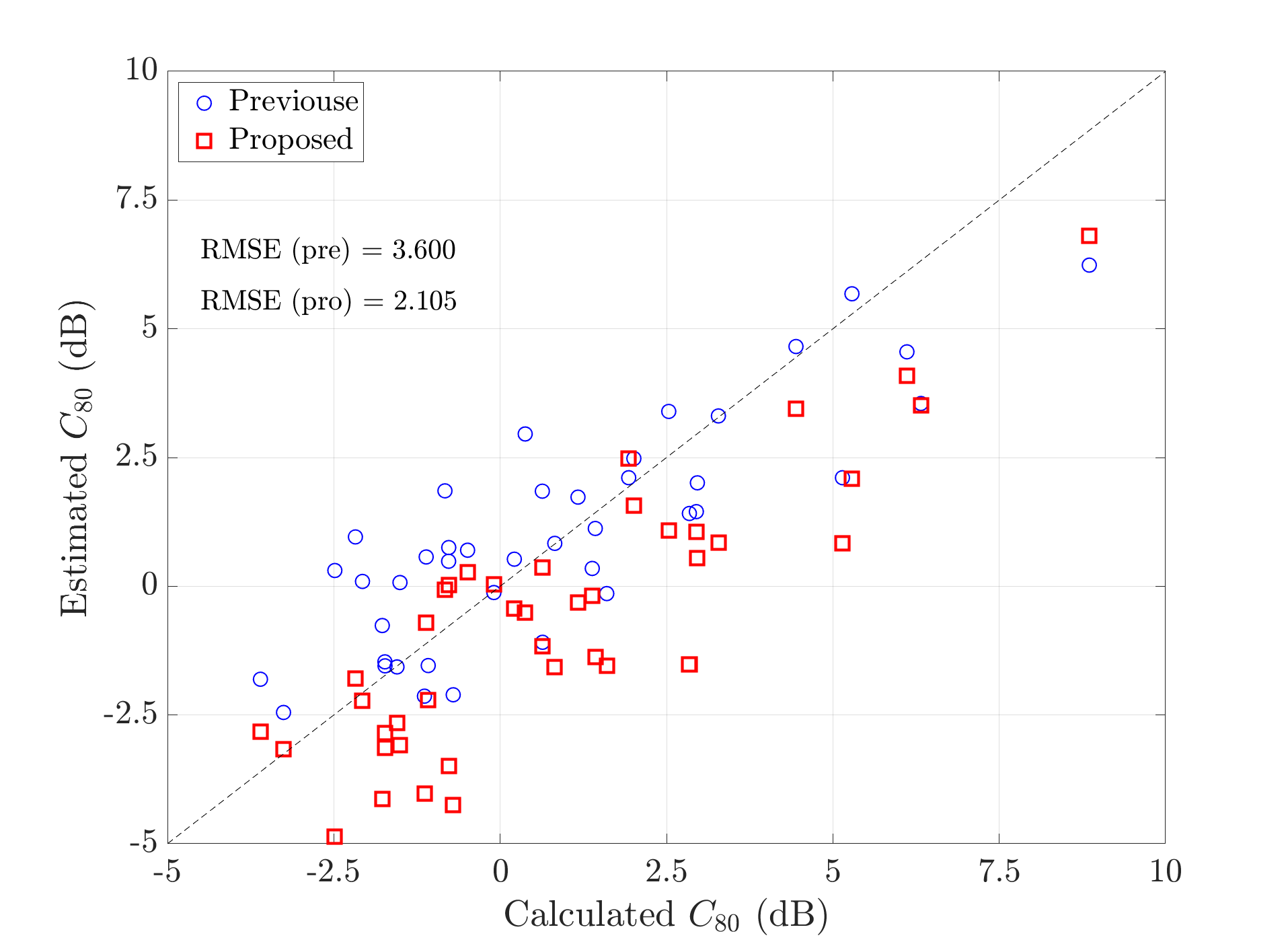}
			\label{fig:C80_SMILE}
		\end{subfigure}
		\hfill
		\begin{subfigure}
			\centering
			\includegraphics[width=0.48\textwidth]{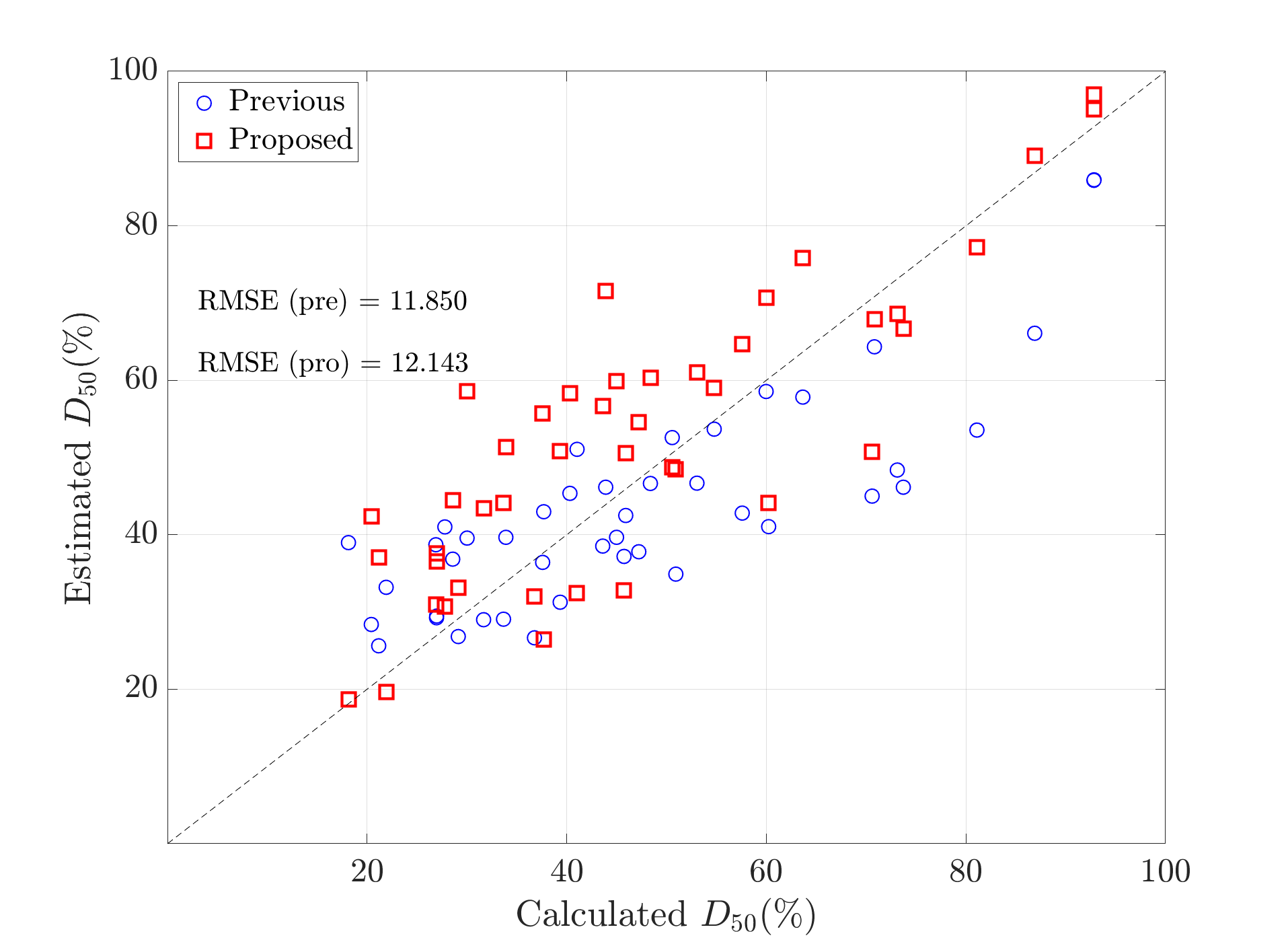}
			\label{fig:D50_SMILE}
		\end{subfigure}
		\hfill
		\begin{subfigure}
			\centering
			\includegraphics[width=0.48\textwidth]{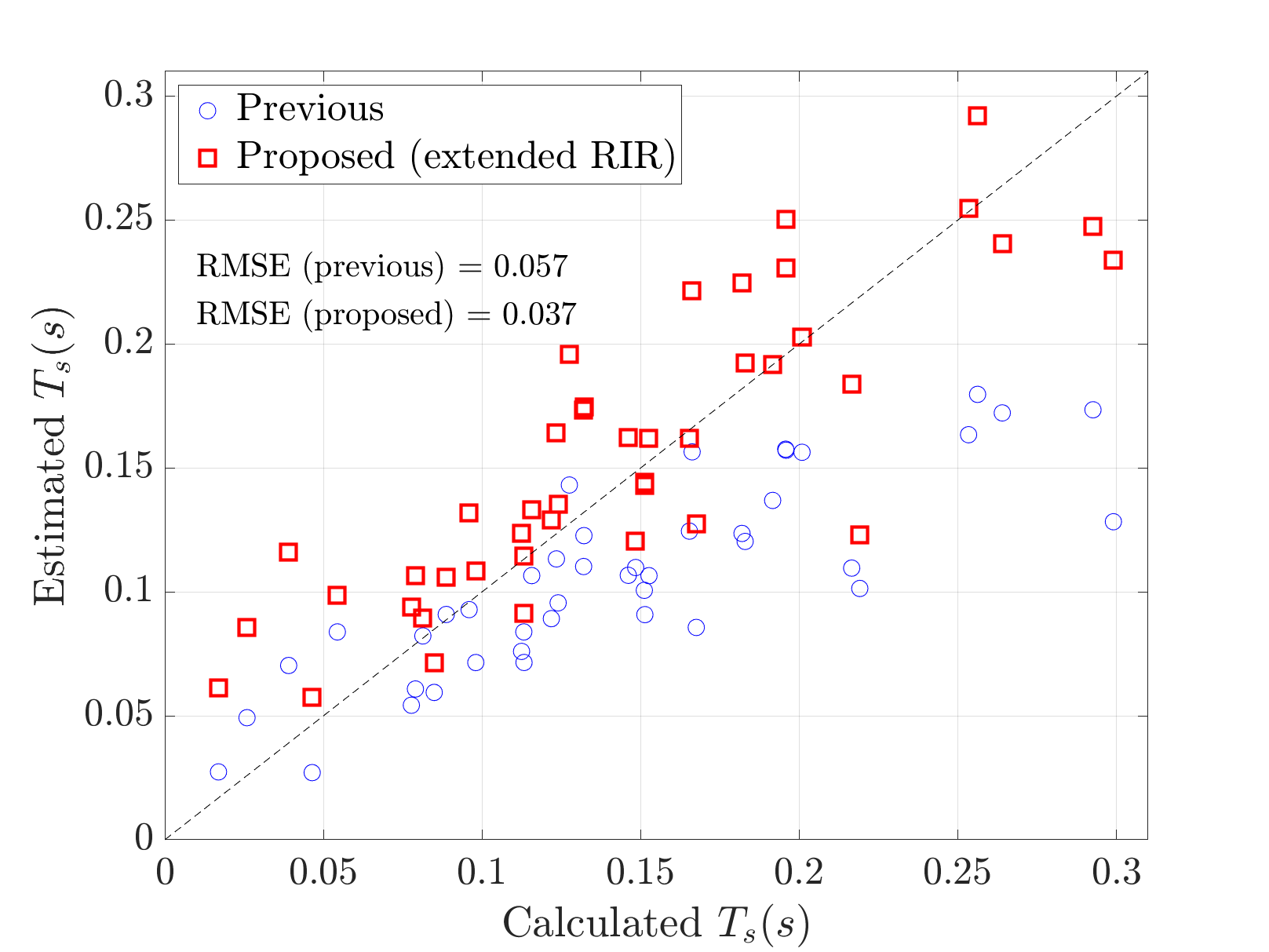}
			\label{fig:Ts_SMILE}
		\end{subfigure}
		\hfill
		\begin{subfigure}
			\centering
			\includegraphics[width=0.48\textwidth]{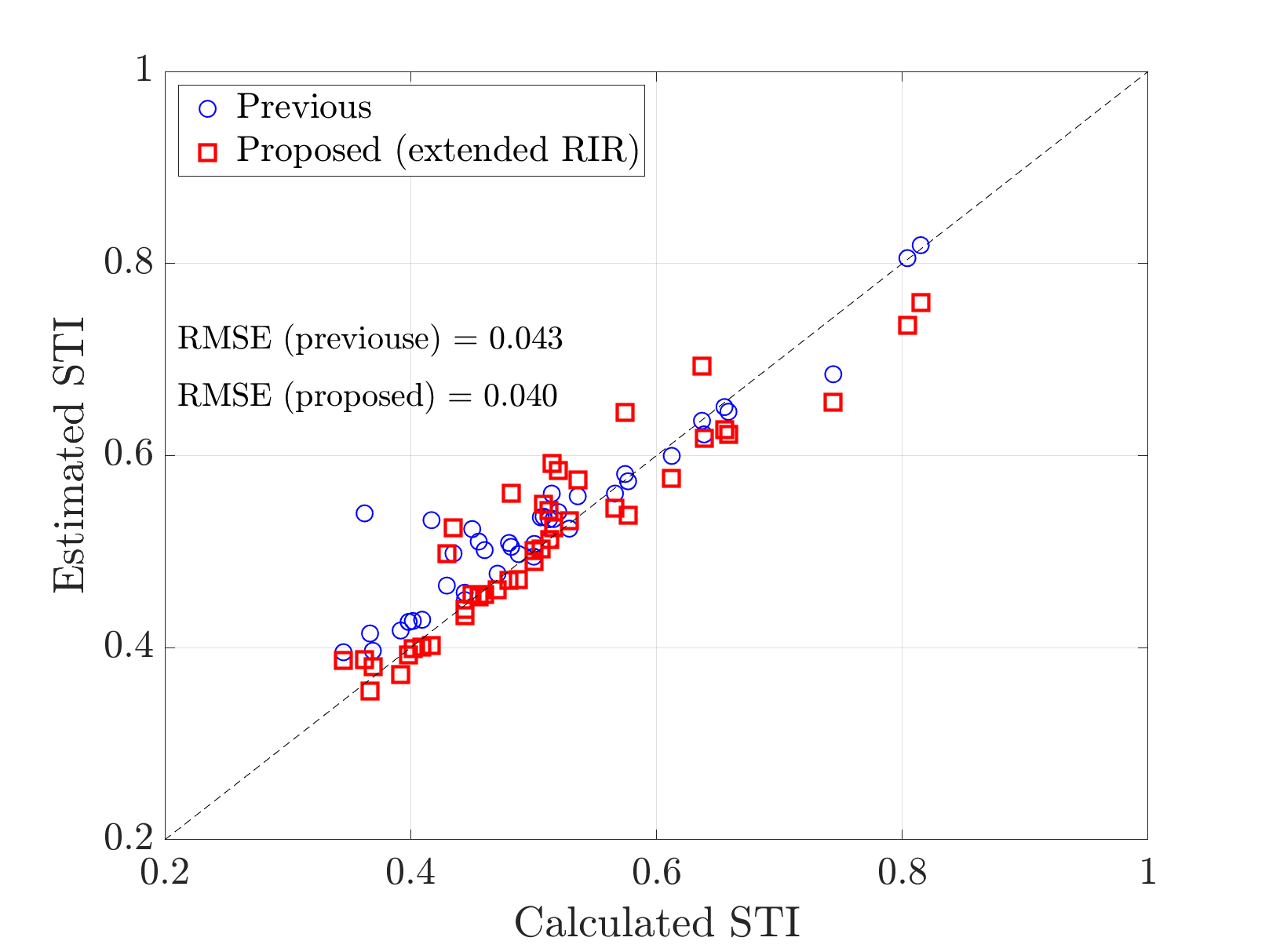}
			\label{fig:STI_SMILE}
		\end{subfigure}
		\centering
		\caption{Estimated results of room acoustic parameters and STI from observed speech signals in reverberant environments: (a) $T_{60}$, (b) EDT, (c) $C_{80}$, (d) $D_{50}$, (e) $T_s$, and (f) STI. The symbol ``o" corresponds to the estimated value from the simulated RIR, ``square" indicates the estimated value from the measured RIR, ``*" indicates the estimated result using the previous method \cite{MU_T60,Me_APSIPA2019}, and the dashed line represents the ground-truth calculated from the RIRs.}
		\label{fig:condition3_SMILEdataset}
	\end{figure*}
	
	\section{Experimental Setup}
	A total of $29,000$ reverberant speech signals with a sampling rate of $16$ kHz were generated from the simulated RIRs convoluted with speech signals. The simulated RIRs are based on Schroeder's RIR model. The reverberation time of the RIRs varies from $0.2$ to $3.0$ s with a step size of $0.1$ s. Each envelope with a different $T_{60}$ was modulated with a different random seed WGN carrier. There are a hundred different WGN carrier seeds. The speech signals were ten short (five-second) Japanese sentences uttered by five men and five women in \citen{ATRspeech}. These reverberant signals were separated into $70\,\%$ for training, and the rest for testing the model. 
	
	Moreover, we evaluated the proposed method to determine whether it can estimate the parameters and STI even though the acoustic characteristics might not follow Schroeder's RIR model. We utilized $43$ realistic RIRs measured from many places in the SMILE dataset \cite{SMILE} for the final evaluation. Here, the accuracy of the estimator can be indicated by using root-mean-square error (RMSE) and correlation coefficient. 
	
	\begin{table}[t]
		\centering
		\caption{Correlation coefficients between the estimated and calculated parameters.}
		\label{tab:rho}
		\begin{tabular}{@{}lllllll@{}}
			
			\toprule
			& $T_{60}$ & EDT & $C_{80}$ & $D_{50}$ & $T_s$ & STI \\ \midrule
			Simulated rooms & 0.996   & 0.996    & 0.992      & 0.994    & 0.996   & 0.997 \\ \midrule
			Real rooms & 0.915   & 0.870    & 0.918      & 0.818    & 0.822   & 0.902\\ \midrule
			\vspace{-5mm}
		\end{tabular}
	\end{table}
	
	\section{Results and Discussion}
	Fig\,\ref{fig:estMTF_fromT60s} shows an example of the approximated MTFs from a speech signal in a simulated room (``o") and real room (``*") where $T_{60}\!=\!0.7$ s. The dashed lines indicate the estimated MTFs, and the solid line is the ground-truth. The averaged $14$ MTFs are derived from the estimated $T_{60}$s in the seven bands. It was found that the shapes of the approximated MTFs were similar to the ground-truths within an RMSE of $0.15$ dB. 
	
	Fig\,\ref{fig:condition3_SMILEdataset} shows the estimated results of the estimated room-acoustic parameters and STI from speech signals in reverberant environments. The symbols ``o" and ``square" correspond to the estimated parameters in the simulated room and realistic room, respectively, where `*" is the value estimated using the previous methods. The horizontal axis indicates the parameter directly calculated from the RIRs, and the vertical axis indicates estimated values. It was found that the results from the simulated rooms were excellent in all parameters. On the other hand, in the real rooms, the results suggested that the proposed method can be used to estimate the five room-acoustic parameters and STI. However, none of the current methods can estimate these parameters simultaneously. We then directly compared only $T_{60}$ and STI with our previous methods \cite{MU_T60,Me_APSIPA2019}. The others were discussed from the results compared with their ground-truths. 
	
	The results of the estimated $T_{60}$ and STI show that the proposed method outperforms the previous methods since it provided significantly lower RMSEs. The estimated $T_{60}$ was improved about $40\%$, and $25\%$ for the STI compared with the previous methods, respectively. For $C_{80}$, $D_{50}$, and $T_s$, the RMSE were $1.66$ dB, $11.85\%$, and $0.06$, respectively. The estimated $C_{80}$ was close to the accuracy from the standard measurement \cite{ISO3381}. However, the estimated $D_{50}$ and $T_s$ have remaining outliers. Those errors might be caused by a mismatch between the RIR model we used and the real RIRs. 
	
	Table\,\ref{tab:rho} shows the correlation coefficients between the estimated parameters and ground-truths. The results show that the proposed method was successful in unseen simulated rooms since the correlation coefficients were close to $1$. For the real rooms, the proposed method has high correlations in all parameters, but the estimated $D_{50}$ and $T_s$ were slightly low.
	
	\section{Conclusion}
	We proposed a blind method for estimating five room-acoustic parameters (i.e., $T_{60}$, EDT, $C_{80}$, $D_{50}$, and $T_s$) and the STI. We leveraged the relationship between a stochastic RIR model and its MTF to estimate $T_{60}$ for seven-octave bands. The proposed scheme estimated $T_{60}$ from the temporal amplitude envelope of an observed signal in each band. The estimated $T_{60}$s were used to approximate the MTF and RIR for deriving of the room acoustic parameters and STI. Simulations were carried out to determine whether the proposed method could estimate the room acoustic parameters and STI from reverberant speech signals even if the RIRs were not the same as Schroeder's RIR model. The experimental results in terms of RMSEs and correlation coefficients showed that the proposed method yielded a better accuracy, compared with the baselines for the STI and $T_{60}$. Also, the estimated EDT, $C_{80}$, $D_{50}$, and $T_s$ were also close to the standard methods.
	
	
\end{document}